# Enhanced spontaneous emission from quantum dots in short photonic crystal waveguides

Thang Ba Hoang[1,*], Johannes Beetz[2], Leonardo Midolo[1], Matthias Skacel[1], Matthias Lermer[2]

Martin Kamp[2], Sven Höfling[2], Laurent Balet[1,3], Nicolas Chauvin[1], Andrea Fiore[1]

[1]*COBRA Research Institute, Eindhoven University of Technology, P.O. Box 513, NL-5600MB Eindhoven, The Netherlands*

[2]*Technische Physik and Wilhelm Conrad Röntgen Research Center for Complex Material Systems, Universität Würzburg, Am Hubland, D-97074 Würzburg, Germany*

[3]*Ecole Polytechnique Fédérale de Lausanne, CH-1015 Lausanne, Switzerland*

We report a study of the quantum dot emission in short photonic crystal waveguides. We observe that the quantum dot photoluminescence intensity and decay rate are strongly enhanced when the emission energy is in resonance with Fabry-Perot cavity modes in the slow-light regime of the dispersion curve. The experimental results are in agreement with previous theoretical predictions and further supported by three-dimensional finite element simulation. Our results show that the combination of slow group velocity and Fabry-Perot cavity resonance provides an avenue to efficiently channel photons from quantum dots into waveguides for integrated quantum photonic applications.

* Author to whom correspondence should be addressed. Electronic mail: t.b.hoang@tue.nl

The generation and control of single photons are critically important for quantum photonic technologies. Semiconductor quantum dots (QDs), proven to be promising candidates as quantum emitters, can provide single and entangled photons on demand.[1] Unlike individual atoms, individual QDs can be embedded in semiconductor devices and can be integrated with complex photonic structures such as photonic crystals, resulting in optimized extraction efficiency and enhanced light-matter interaction. For example, when a QD emits a photon into a photonic crystal cavity or waveguide, the QD spontaneous emission can be enhanced through the Purcell effect.[2]

In earlier works,[3-6] a system was proposed where single quantum dots emit single photons into the slow-light mode of a photonic crystal waveguide (PhC WG). This "photon-gun on a chip" scheme has also been experimentally demonstrated.[7-9] Some clear advantages of such a system are: i)- the slow-group velocity mode presents an increased local density of states and therefore increased coupling to the emitter (Purcell effect); ii)- the coupling of the QDs to a PhC WG mode is easy to achieve due to the large WG volume and relatively broad spectral resonant range.[8] This is particularly important because, for example when a QD is placed in a photonic crystal cavity with a small volume, the precise alignment (spectral and spatial) of the QD position with the cavity is challenging. On the other hand the high quality factors and typical Purcell enhancements (such as observed in photonic crystal cavities) are in practice difficult to achieve in long PhC WGs due to unavoidable disorder effects.[10-14]

In this work we examine a mixed approach where a combination of the slow-light effect with cavity enhancement is achieved by embedding semiconductor quantum dots in short (10-25 μm) PhC WGs. In these short PhC WGs the localization due to disorder is not expected to occur since the PhC WG length becomes smaller than the photon localization length. Earlier Rao and Hughes[4] have also theoretically examined a similar structure and shown that a large coupling efficiency and robustness with respect to disorder can be obtained. The interference effect due to the reflection at the two WG end facets results in the formation of Fabry-Perot (FP) modes in the slow-light part of the dispersion curve. We observe that the QD spontaneous emission rate is enhanced when its emission energy is in resonance with these FP cavity modes resulting in large coupling efficiency.

We first present the simulation result of a W1 PhC WG (PhC membrane with one missing air hole row) which is similar to those studied in the optical characterizations. The computation principle of the spontaneous emission enhancement rate was adapted from Y. Xu et al.[15] However, here we employ the 3D Finite Element Method instead, and the following parameters were used: lattice spacing $a = 306$ nm, filling factor $f = 31\%$, GaAs membrane thickness $h = 320$ nm and the length of the PhC WG $L = 9$ μm. A Y-polarized source dipole is placed at the center of the WG. In Fig. 1a we show a map which indicates the distribution of the field $|Ey|^2$ (in-plane polarization and perpendicular to the WG axis) at wavelength 1280.45 nm. Fig. 1c displays the field $|Ey|^2$ (left axis, probed at the location marked by a crossed circle $\otimes$ in Fig. 1a) and the spontaneous emission rate normalized to the spontaneous emission rate in bulk GaAs (right axis) as functions of wavelength. We observe clear enhancements of the electric field component $|Ey|^2$ as the frequency approaches the bottom of the dispersion curve of the even mode in the W1 WG near ~1280.45 nm. The pronounced peaks appearing near the cut-off frequency (1280.45 nm) are the FP resonances produced by the two WG end-facets. More importantly, when calculating the spontaneous emission rate we observe similar enhancements due to resonance with the FP cavity modes in the slow-light region. We note that a similar combination of slow-light dispersion and spatial confinement has been theoretically proposed for 2D PhCs by Bordas et al.[16] and Nedel et al.[17] A similar QD spontaneous emission rate dues to combination of slow-light and FP resonances was also predicted in ref. 4. We can compare our simulation results with the simple prediction given by the Purcell factor $F_p = (3/4\pi^2) \times Q/V_{eff}$ which usually applies to cavities. From the electric field simulation we find at the resonance wavelength $\lambda = 1280.45\,nm$ an effective mode volume of $V_{eff} \sim 4 \times (\lambda/n)^3$ ($n$ being the effective index of the slab) and a quality factor $Q \approx 16500$, giving a value $F_p = 310$ which is close to the value of the spontaneous emission enhancement rate presented in Fig 1c. We therefore conclude that the usual treatment of the Purcell enhancement in term of $V_{eff}$ and Q applies reasonably well to these slow group velocity enhanced FP modes. We note that the main role of the low group velocity is to increase the photon lifetime in the WG and therefore increase Q of the FP mode, as it is clear from the increased Q factor for modes in the slow-light region.

For the fabrication of the PhC WGs, a 1.5 μm thick AlGaAs sacrificial layer and a 320 nm thick GaAs slab were grown on a GaAs substrate by molecular beam epitaxy. The slab contains an InAs QD layer (a few QDs/μm$^2$) emitting near 1300 nm. The PhC hole pattern was defined by electron beam lithography and transferred from the resist into an underlying SiO2 layer by CHF$_3$/Ar reactive ion etching. The SiO$_2$ layer serves as etch mask for a subsequent Ar/Cl$_2$ electron cyclotron resonance reactive ion etching of the slab. The W1 PhC WGs were formed by leaving a row of holes un-patterned. Afterwards, the sample was cleaved and exposed to hydrofluoric acid in order to both under-etch the PhC and remove the remaining SiO$_2$ mask layer. Cleaving after etching results in worse facet smoothness compared to our sample shown in Fig. 1b. The WG length *L* was varied between 10 and 25 μm. Such a WG length is long enough to produce the slow light effect and yet short enough to minimize the effect of disorder due to fabrication imperfections. The PhC lattice spacing *a* was varied from 300 nm to 340 nm (5 nm step) in order for the slow-light frequencies of the PhC WGs to be in resonance with the QD ground state emission at both low and room temperatures (~1260 nm and 1350nm, respectively). In the optical measurements, we used a specially designed low-temperature micro-photoluminescence (µ-PL) set-up where two long working distance objectives were used to excite the QDs (from top) and collect PL (from top and at the WG side) through a cubic cryostat window. The PL emission from single QDs can be collected from the top and from the WG facet simultaneously. After being collected by the objectives, emitted photons are coupled into single mode optical fibers, dispersed by a f=1m spectrometer and either detected by a charge-coupled device (CCD) camera or by a superconducting single photon detector (SSPD) for time-resolved measurements.[18]

We first explore the slow group velocity band by plotting in Fig. 2b two PL spectra which were collected from the top (dashed, blue) and the side (solid, red) with the pumped laser beam focused close to the center of a PhC WG (*L*=17 µm) at room temperature. For an easy comparison, we show in Fig. 2a the band diagram of a W1 PhC WG which was calculated by the MIT Photonic Bands (MPB) software,[19] where two guided modes under the light-line (dashed line) are observed, corresponding to the fundamental (even, red squares) and first-order (odd, blue squares) modes. The two pronounced features clearly observed from the top and side facet shown in Fig. 2b are originated from these two modes. When collecting PL from the WG top, we mainly observe a peak at a wavelength of 1220 nm ($a/\lambda \sim 0.278$), corresponding to the slow-group velocity regions (at k=0 and k=π/a) of the dispersion curve of the odd mode.[20] This mode, polarized along the WG axis, does not radiate from the WG facet and is therefore not visible in the PL collected from the WG side. On the other hand, when collecting from the WG side facet, one can clearly see a series of sharp, pronounced peaks on top of a broad peak at around 1315-1330 nm ($a/\lambda \sim 0.252-0.257$) in the PL spectrum. Both the broad emission and pronounced sharp peaks are polarized in the plane, perpendicularly to the WG axis. The broad peak emission is the enhancement of the QD ground state emission which is in resonance with the slow-group velocity region of the dispersion relation of the even mode. The sharp peaks (linewidth decreasing from 1.8 nm to 0.2 nm at longer wavelength side) are FP resonant modes due to the reflection at the two ends of the PhC WG which occur in the slow-group velocity regime. We note that in contrast to the localization by randomness as in the case of long PhC WGs,[10-14] here we have a localization which can be controlled by the PhC WG termination. The interesting differences between the spectra measured from the waveguide side and from the top indicate preferential emission into different modes depending on the wavelength. The FP resonances in the slow-light part of the dispersion curve, as we will present in detail below, may provide sufficient enhancement of the optical density of states to achieve efficient photon emission.

We will now investigate the structural details of the FP resonances in the slow group-velocity regime. Fig. 3a shows the PL spectrum collected from the side facet of a PhC WG with lattice constant of 340 nm and length $L=17$ μm. As one can see, the wavelength spacing between the resonant peaks is not constant and, in fact, it decreases with increasing emission wavelength. From this wavelength dependence we can extract the group index using $n_g = \frac{\lambda^2}{2L\Delta\lambda}$, $\lambda$ the emission wavelength, and $\Delta\lambda$ the distance between neighboring resonant peaks. The result plotted in Fig. 3b shows a clear signature of slow group velocity. We further estimate the quality Q-factor $Q = \frac{\lambda}{\delta\lambda}$ (where $\delta\lambda$ the peak's linewidth) of the FP resonances in Fig. 3a and the result is shown in Fig. 3c. As predicted by the model (Fig. 1), we observe a significant increase of the Q-factor as the emission frequencies approach the cut-off frequency near 1344.5 nm. The increase of the Q-factor implies that the time photons spend in the WG is longer, as a consequence of the smaller group velocity, assuming that the main loss comes from the transmission at WG facets.

As mentioned above, an advantage of coupling the QD emissions to the slow-group velocity waveguide mode is the enhancement of the QD spontaneous emission due to the increase of the local density of state in the slow-light part of the dispersion curve. In Fig. 4b we show a high resolution micro-PL spectrum measured from a side facet of a PhC WG ($L$=15 μm) at low laser excitation power and low temperature (10K). Several FP modes are observed (which are clearly visible when measured at higher laser excitation power as shown in Fig. 4a) together with narrower (multi) excitonic lines, some of which are resonant with a FP mode. The nature of the excitonic lines was verified by changing the temperature and laser excitation power (not shown here). In Fig. 4c we plot the corresponding decay times measured at different spectral positions (by using a spectrometer as a spectral filter with bandwidth ~ 0.5 nm). Overall, we see that the decay time decreases from 1.6 ns, when the QD emission is away from resonance, down to ~ 0.8 ns as the emission energy approaches the band-edge. We can also clearly see in Fig. 4c that at each FP resonance frequency, the decay time is reduced as compared to its neighboring frequencies. This is the result of the local enhancement due to the combined effect of the slow-light with the FP resonances, as predicted from the calculations shown in Fig. 1. The decreasing of the decay time, which reflects well the findings from Fig 3, is a strong evidence of the increasing of the local density of state in the slow light regime as compared to the rest of the dispersion curve. The Purcell factor $Fp$ is ~1.7 for the peaks at 1266.4 nm and 1267.5 nm taking into account that the measured lifetime of these dots in bulk GaAs is 1.3-1.4 ns. Experimentally observed spontaneous emission enhancement values (i.e the decreasing of the decay time) are much lower than the calculated ones because of limited spectral and spatial matching of single QDs with FP modes in the slow-light regime. In principle a large Purcell factor can be obtained by finding a single QD spectrally resonant with the FP peak closest to the bottom of the dispersion curve where a high Q and high spontaneous emission rate present. However here we like to point out that a more interesting quantity is the $\beta$-factor which reveals the coupling efficiency of the QD emission into the slow light mode. The highest $\beta$-factor for these FP modes can be estimated as $\beta = 1 - \frac{\tau_{on}}{\tau_{PhC}} \sim 80\%$. Here $\tau_{on} = 0.8\,\text{ns}$ is the decay time measured when the QD emission is on-resonance with the modes (for example the peaks at 1266.4 nm and 1267.5 nm in Fig. 4) and $\tau_{PhC} \sim 4\,\text{ns}$ is the decay time measured in similar cavities from QDs emitting at longer wavelengths, i.e. in the frequency region between the PhC band-edge and the bottom of the dispersion curve of the even mode. The time $\tau_{PhC}$ thus quantifies the effect of emission rate into leaky modes and nonradiative recombination. We note that a high value of the $\beta$-factor was also previously predicted (ref.5: 95%, ref.6: > 85%) and demonstrated (ref.8: 89%). The high $\beta$ value indicates that even with limited spectral and spatial mismatching, we can still have an efficient coupling between the QD emission and the WG slow light mode. This is because of the broad spectral range of the slow light mode (~ 15-20 nm in the present study) and the large PhC WG mode volume.

In conclusion we have shown that by combining the slow light effect and FP cavity enhancement in short PhC WGs, one can efficiently extract photons from semiconductor quantum dots into a single WG mode. The results of our study provide an avenue to the manipulation and extraction of photons on a chip and are critically important steps on the road toward future integrated photonic circuits.

We acknowledge financial support from the European Commission within the FP7 project QUANTIP (project n. 244026) and from the Dutch Technology Foundation STW, applied science division of NWO and the Technology Program of the Ministry of Economic Affairs. We thank Prof. Peter Lodahl for the fruitful discussions.

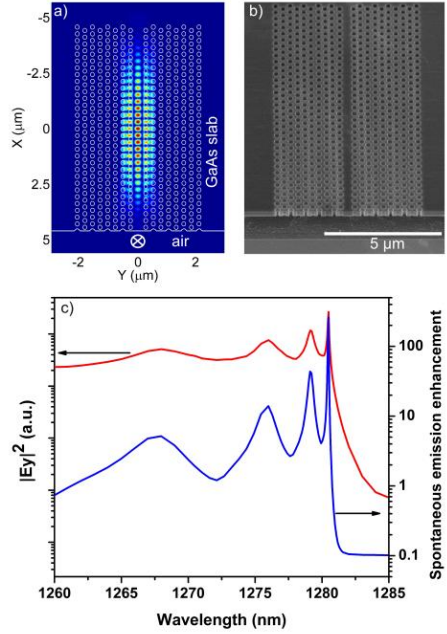

FIG. 1. a) A simulation shows the top view map of a slab with the field $|E_y|^2$ distribution corresponding to the peak at 1280.45 nm shown in (c) b) a scanning electron microscopy image of a fabricated PhC WG membrane and c) in-plane field vector $|E_y|^2$ (left axis, probed at a location marked by $\otimes$ in a)) and the spontaneous emission enhancement rate (right axis).

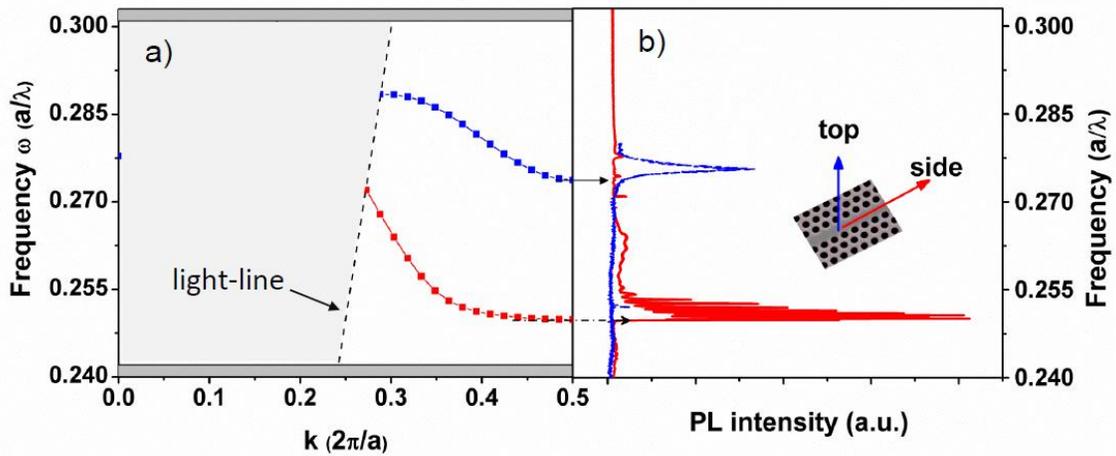

FIG. 2. a) Band structure of a W1 PhC WG calculated by MPB.[19] Two guided modes under the light-line (dashed line) were observed. Two shaded areas indicate the photonic band gap boundaries b) PL emission spectra of a W1 PhC WG which are being collected from the WG's top and its cleaved facet. Arrows indicate corresponding emission modes.

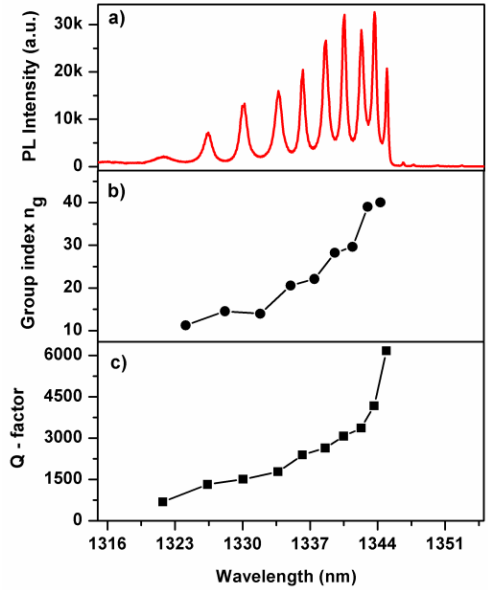

FIG. 3. a) A close look at the FP mode structure b) the group index $n_g$ as a function of the emission wavelength and c) the Q-factor of the FP resonances as a function of the emission wavelength.

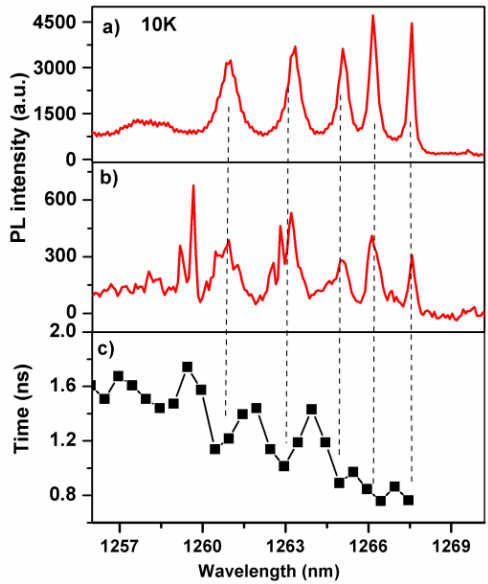

FIG. 4. a) Side collected PL spectrum of a PhC WG under high laser excitation power – clear FP resonant peaks were observed. b) same WG but under lower laser excitation power – visible single QD emissions coupled to the FP modes are observed. c) decay times measured at different emission wavelengths.